\documentclass[a4paper,11pt]{article}
\pdfoutput=1 % if your are submitting a pdflatex (i.e. if you have
\usepackage{jinstpub_for_arxiv}
\usepackage[utf8]{inputenc}
\usepackage{siunitx}

\title{\boldmath MuPix7 -- A fast monolithic HV-CMOS pixel chip for Mu3e}

\author[a]{H.~Augustin,}
\author[b]{N.~Berger,}
\author[a]{S.~Dittmeier,}
\author[a]{J.~Hammerich,}
\author[b]{U. Hartenstein,}
\author[b]{Q.~Huang,}
\author[a]{L.~Huth,}
\author[a]{D.~Immig,}
\author[b]{A.~Kozlinskiy,}
\author[a,1]{F.~Meier~Aeschbacher,\note{Corresponding author.}}
\author[c]{I.~Perić,}
\author[a]{A.-K. Perrevoort,}
\author[a]{A.~Schöning,}
\author[a,2]{S. Shrestha,\note{Now at Middle Tennessee State University}}
\author[b]{I. Sorokin,}
\author[b]{A. Tyukin,}
\author[b]{D.~vom~Bruch,}
\author[b]{F.~Wauters,}
\author[a]{D.~Wiedner,}
\author[b]{M. Zimmermann}

\affiliation[a]{Universität Heidelberg, Physikalisches Institut, Im Neuenheimer Feld 226, 69120 Heidelberg, Germany}
\affiliation[b]{Institut für Kernphysik, Johann-Joachim-Becherweg 45, Johannes Gutenberg-Universität Mainz, 55128 Mainz, Germany}
\affiliation[c]{Institut für Prozessdatenverarbeitung und Elektronik, KIT, Hermann-von-Helmholtz-Platz 1, 76344 Eggenstein-Leopoldshafen, Germany}

\emailAdd{meier@physi.uni-heidelberg.de}

\abstract{The MuPix7 chip is a monolithic HV-CMOS pixel chip, thinned down to \SI{50}{\micro\metre}. It provides continuous self-triggered, non-shuttered readout at rates up to \SI{30}{\mega hits/chip} of $\SI[parse-numbers = false]{3 \times 3}{\milli\metre\squared}$ active area and a pixel size of $\SI[parse-numbers = false]{103 \times 80}{\micro\metre\squared}$. The hit efficiency depends on the chosen working point. Settings with a power consumption of \SI{300}{\milli\watt / \centi\metre\squared} allow for a hit efficiency $>99.5\%$. A time resolution of \SI{14.2}{\nano\second} (Gaussian sigma) is achieved. Latest results from 2016 test beam campaigns are shown.}

\arxivnumber{1234.56789} % only if you have one

\collaboration[c]{on behalf of the Mu3e collaboration}

\proceeding{8$^{\text{th}}$ International Workshop on Semiconductor Pixel Detectors for Particles and Imaging\\
  September 6--9, 2016\\
  Sestri Levante, Italy}

\begin{document}
\graphicspath{{./figures/}}
\maketitle
\flushbottom

\newcommand{\todo}{\textcolor{red}{TODO: }}
%==================================================
\section{Introduction}
\label{sec:intro}
CMOS pixel detectors are successfully used for tracking detectors in particle physics since years. They provide a cost-effective alternative to hybrid designs, come with small pixel sizes and can be thinned down to \SI{50}{\micro\metre}. Their use in high-rate environments was limited by readout deadtime (shuttered readout) and charge collection limits (thin depletion zones, electron diffusion). The MuPix7 is a pixelated silicon detector made with industry-standard high-voltage CMOS technology, based on the principles described in~\cite{peric}. This allows to apply a bias voltage of up to \SI{-85}{\volt}, hence electron drift dominates. The chip has an array of $32 \times 40$~pixel cells, sized $\SI[parse-numbers = false]{103 \times 80}{\micro\metre\squared}$ each. The charge-sensitive amplifier and the line driver are on top of the (deep-implant) sensor diode in the pixel matrix. The signal is transmitted via a single-ended transmission line to the mirroring digital pixel in the periphery which contains an amplifier, a tuneable comparator, and a time-stamp generator. This design choice protects the analog part from digital noise by separation and results in a fast, zero-suppressed readout that operates continuously without a trigger. Any rate limit comes from shaping time in the pixel cell and the data readout capabilities. The need for many connections from the active array to the periphery consumes space and imposes limits on the integration density, leading to somewhat larger pixels compared to other monolithic CMOS pixel chips. A more detailed description of the MuPix7 chip can be found elsewhere~\cite{mpx7}. This report provides new results to the ones reported therein.

The results shown below have all been obtained in 4-plane telescope setups~\cite{tele} made with thinned MuPix7 chips, unless where noted differently. Tracks were established from three hits and extrapolated to the device under test (DUT). Unless where stated differently, a matching hit on the DUT has to be within a radius of \SI{800}{\micro\metre} around the extrapolated track center and within a time window of \SI{\pm 48}{\nano\second} with respect to the track time.

%==================================================
\section{Efficiency studies}
\label{sec:effstudies}
\begin{figure}[htbp]
\centering % \begin{center}/\end{center} takes some additional vertical space
\includegraphics[width=.48\textwidth]{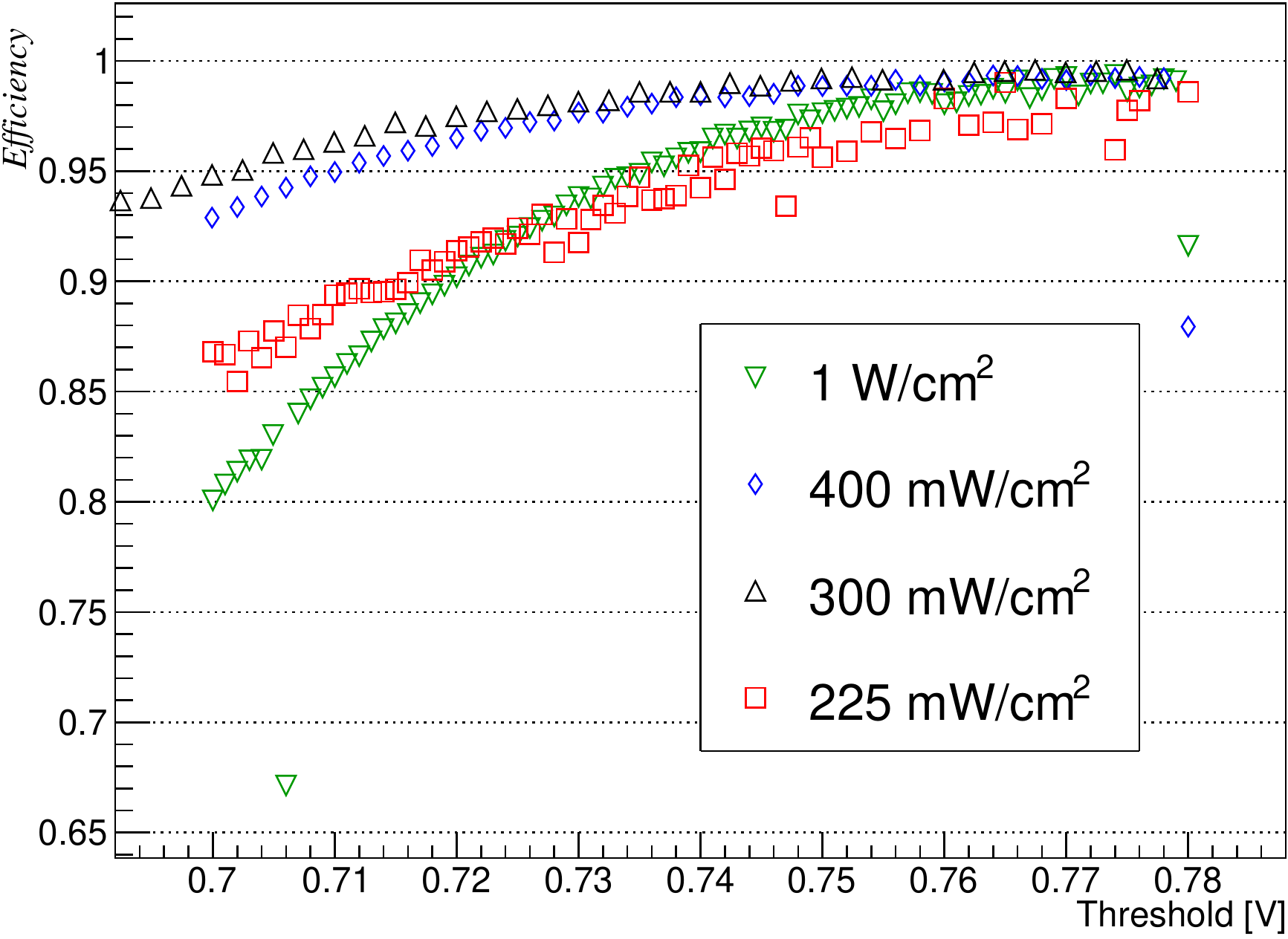}
\quad
\includegraphics[width=.48\textwidth]{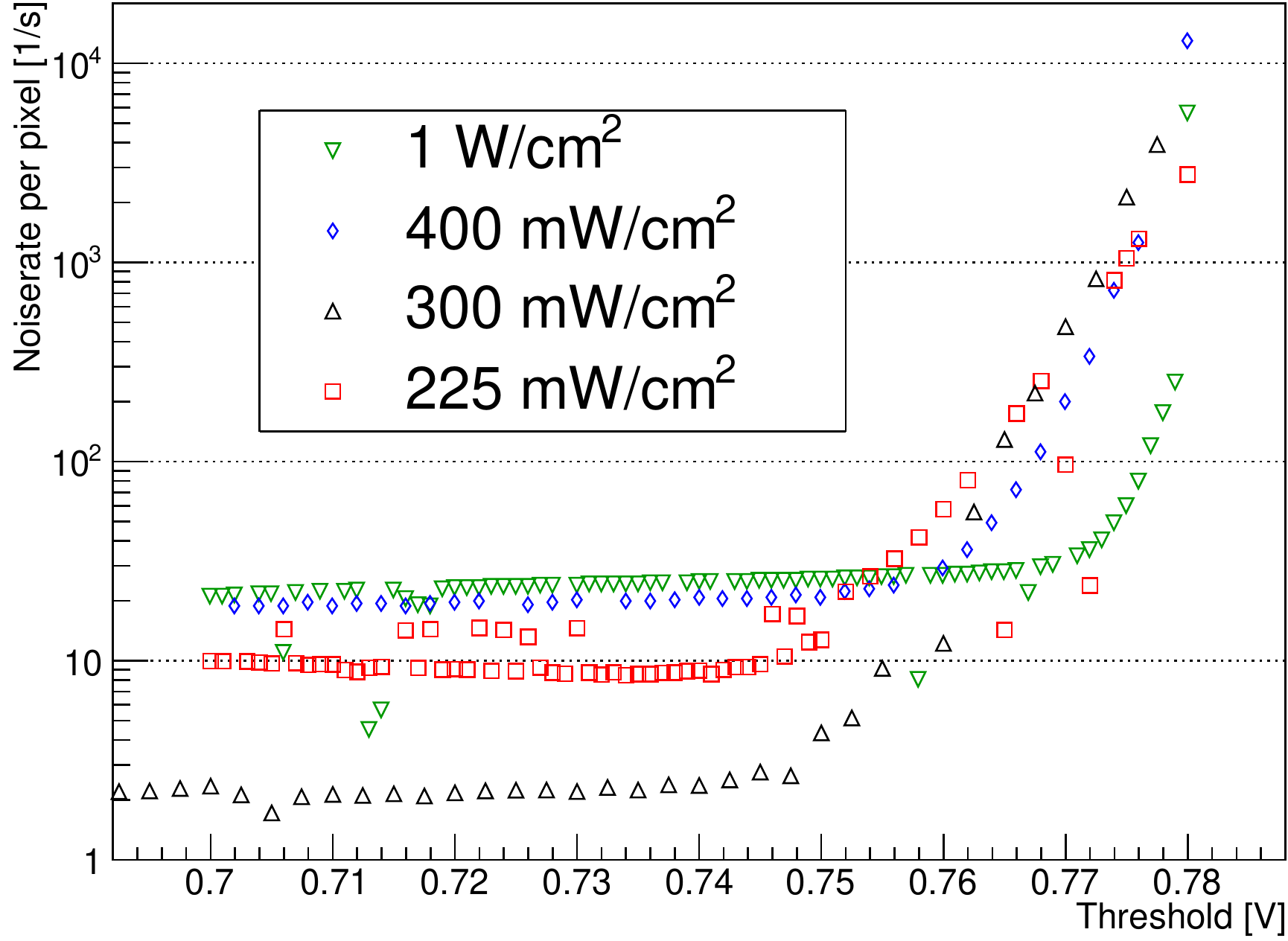}
\caption{\label{fig:effPower} Efficiency (left) and noise (right) as function of threshold for different power settings (higher voltage corresponds to lower comparator threshold). Measured at PSI $\pi$M1, mixed $\pi^+, \mu^+, e^+$ beam at \SI{224}{\MeV} momentum setting.}
\end{figure}

The MuPix7 chip performance is controlled by a number of voltage settings, steered by internal DACs. These settings can be optimised for a variety of performance targets, which are mainly efficiency and time resolution, both at the expense of noise level and power consumption. The power consumption was measured on the external power lines. The efficiency and noise level as function of threshold was measured at four choices of settings and are shown in Fig.~\ref{fig:effPower}. The chip can be operated with an efficiency well above 99.5\% at power ratings equal or below \SI{400}{\milli\watt/\centi\metre\squared}.

\begin{figure}[htbp]
\centering % \begin{center}/\end{center} takes some additional vertical space
\includegraphics[width=.45\textwidth]{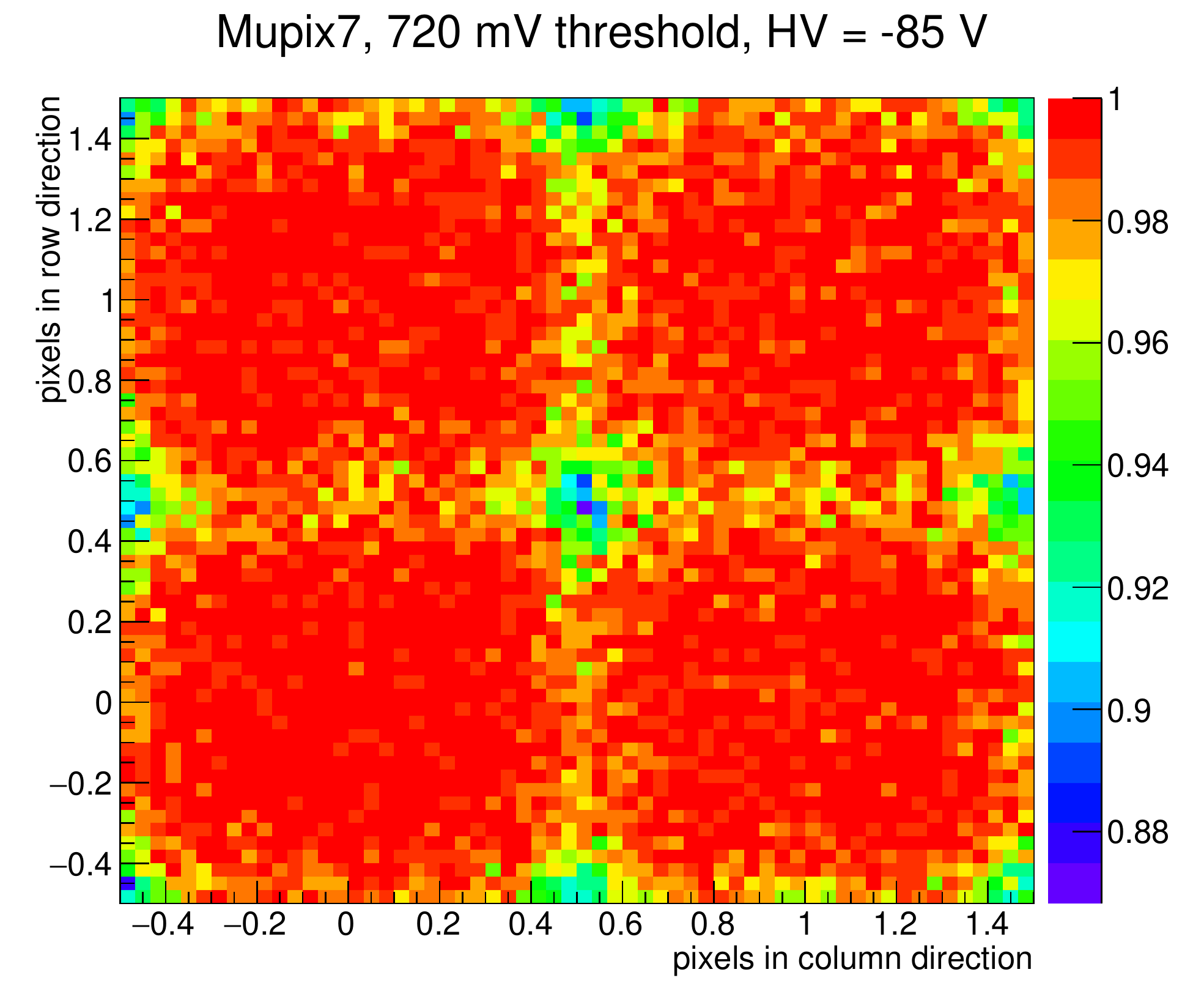}
\includegraphics[width=.47\textwidth]{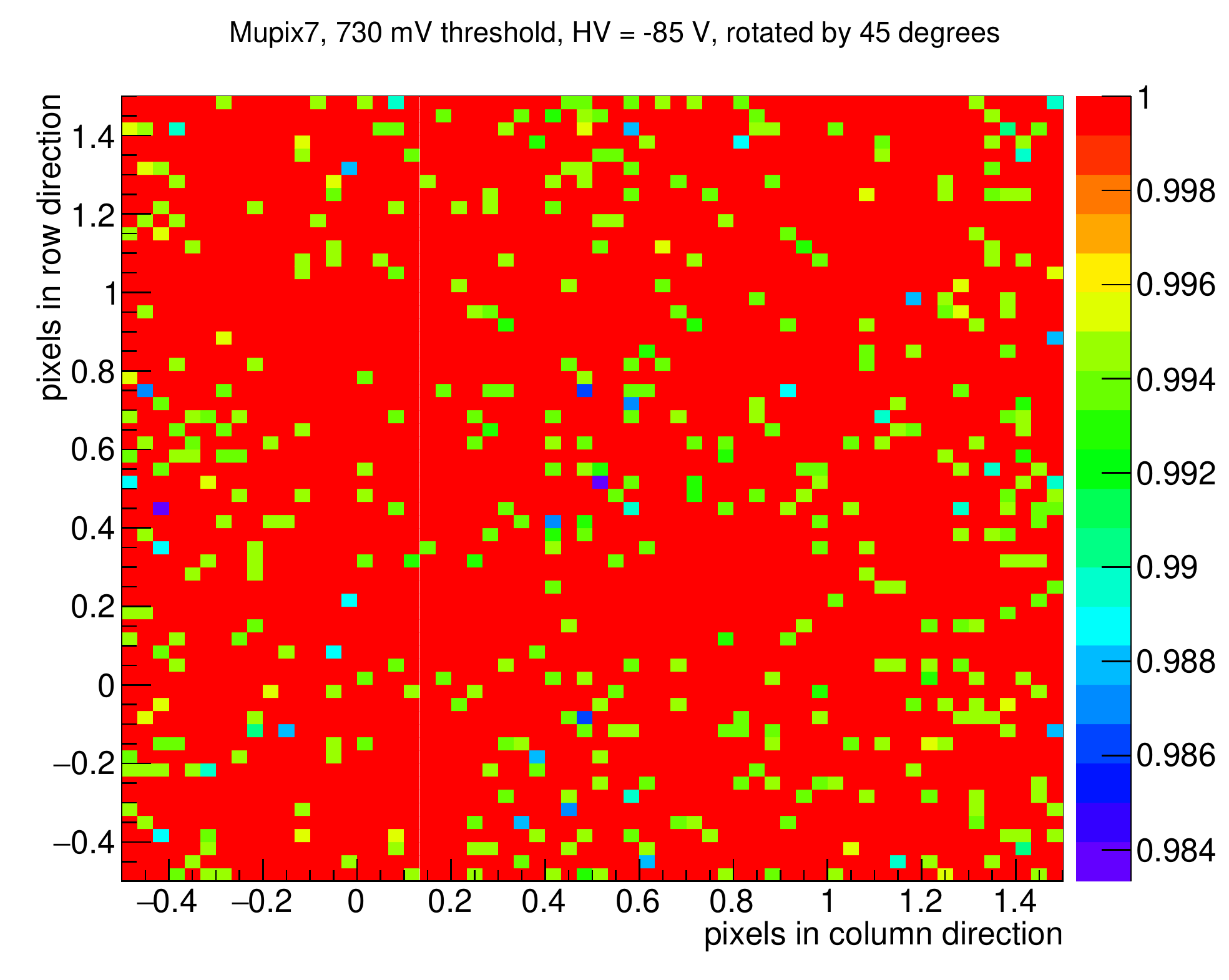}
\caption{\label{fig:effSubpixel} Efficiency map with sub-pixel resolution. Left: Sensor perpendicular to beam axis. Right: Sensor inclined by \SI{45}{\degree} w.r.t.~beam axis. The chip is intentionally operated at higher threshold to force lower overall efficiency for effect enhancement. Hits of all pixels of the chip superimposed on $2\times2$ pixels to enhance the number of entries per bin. Units are pixel size. Measured at DESY using electrons at \SI{4}{\GeV}.}
\end{figure}

Using the EUDET telescope at DESY~\cite{desytele}, we made use of the excellent spatial resolution of the MIMOSA26 chips (about \SI{4}{\micro\metre} at the chosen configuration, measured using track to hit residuals, perpendicular beam incidence) and studied the efficiency of the MuPix7 with sub-pixel resolution by extrapolating the track from the EUDET planes to the MuPix7 DUT. To enhance effects, the threshold of the MuPix7 was detuned targeting a lower overall efficiency. In the MuPix design, charge sharing between neighbouring pixels is a minor effect on the percent level. This is a consequence of the small depletion zone (about \SI{15}{\micro\metre}) compared to the pixel cell size and the fast charge collection. Raising the threshold (i.e.~lowering the threshold voltage) should therefore lower the efficiency for hits at the edges and corners. The results in Fig.~\ref{fig:effSubpixel} (left) clearly show this effect. This can be compensated by turning the sensor by \SI{45}{\degree} with respect to the beam axis, which increases the effective length of the depletion zone by $\sqrt{2}$, also shown in Fig.~\ref{fig:effSubpixel} (right).

%==================================================
\section{Time resolution}
\label{sec:timeres}
\begin{figure}[htbp]
\centering % \begin{center}/\end{center} takes some additional vertical space
\includegraphics[width=.7\textwidth]{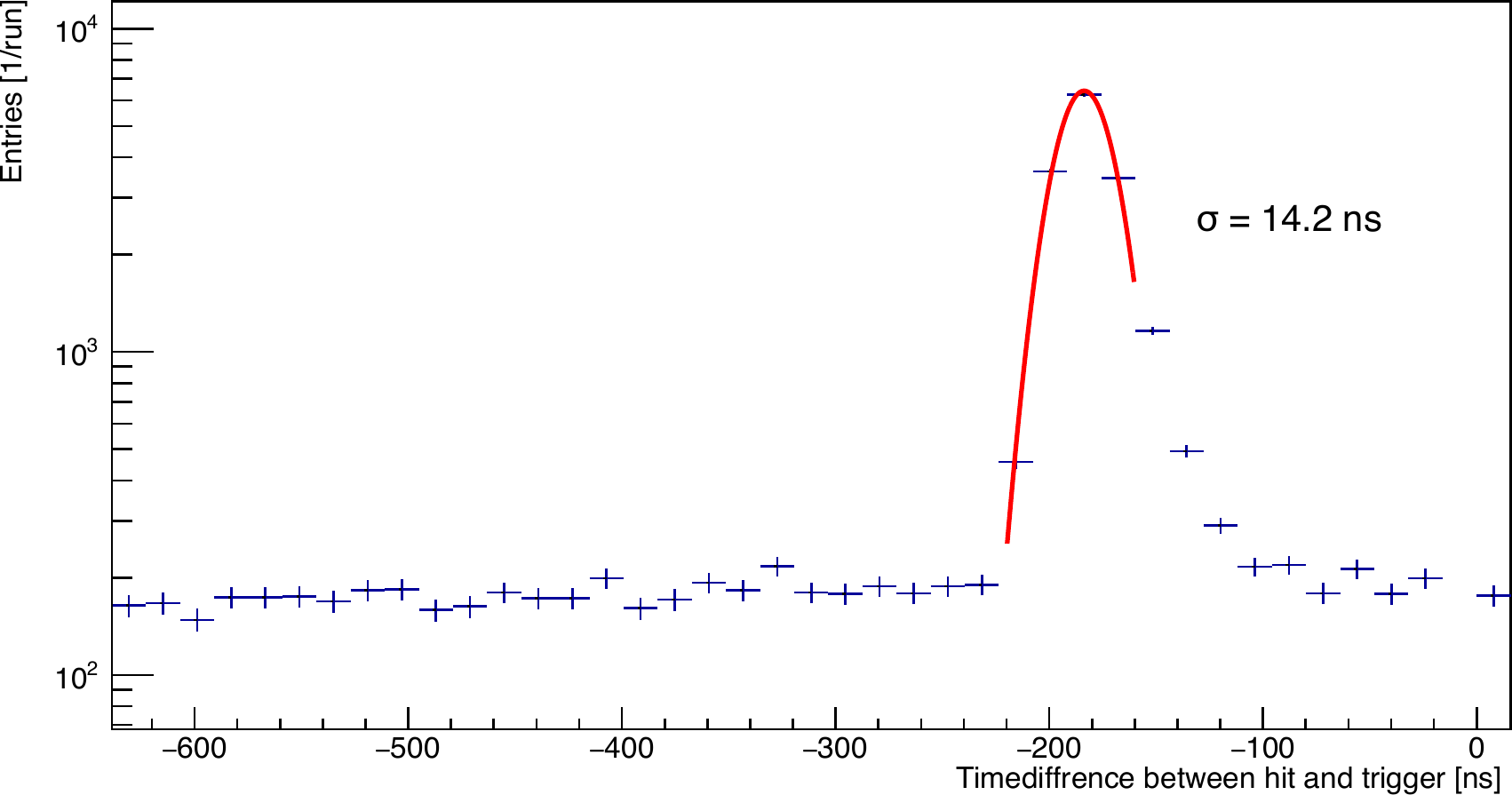}
\caption{\label{fig:timing} Timing resolution measured using settings with a power consumption of \SI{300}{\watt/\centi\metre\squared}. Shown is the difference in time between the hit in the DUT and the scintillators. Bin size is \SI{16}{\nano\second}. Measured at DESY using electrons at \SI{4}{\GeV}.}
\end{figure}

In a previous publication~\cite{mpx7} we reported a timing resolution of \SI{11}{\nano\second} (expressed as Gaussian $\sigma$), measured using the settings corresponding to a power consumption of \SI{1000}{\milli\watt/\centi\metre\squared}. We repeated that measurement using the same setup, where the telescope was amended with scintillators (time resolution about \SI{1}{\nano\second}) for obtaining the timing reference. The radius for a matching hit was set to a tight value of \SI{120}{\micro\metre} and the $\chi^2$ per degree of freedom of the track fit was required to be less than~5. Operating the chip at \SI{300}{\milli\watt/\centi\metre\squared}, a timing resolution of \SI{14.2}{\nano\second} was measured (averaged over all pixels of the DUT), see Fig.~\ref{fig:timing}. This result meets the Mu3e requirements of a timing resolution better than \SI{20}{\nano\second}.

%==================================================
\section{Crosstalk}
\label{sec:xtalk}
\begin{figure}[htbp]
\centering % \begin{center}/\end{center} takes some additional vertical space
\includegraphics[width=.7\textwidth]{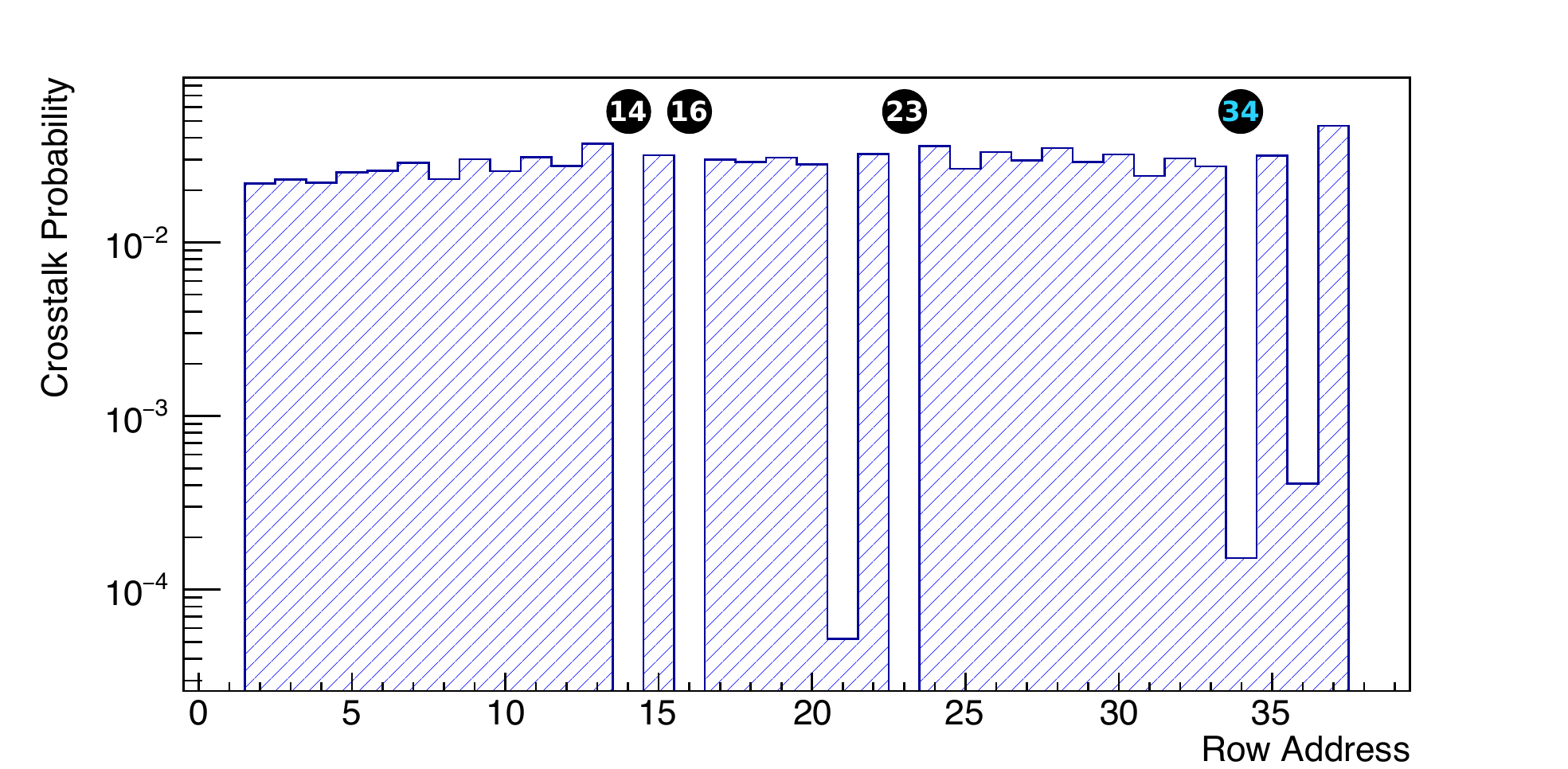}
\quad
\includegraphics[width=.6\textwidth]{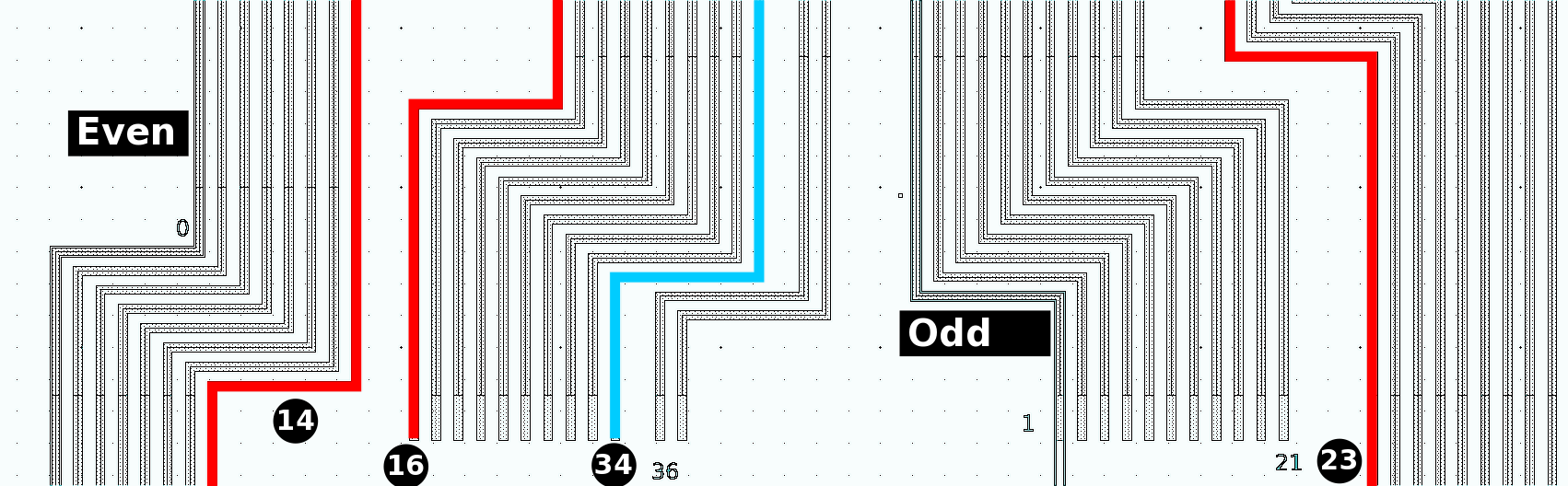}
\caption{\label{fig:xtalk} Top: Crosstalk probability. See main text for explanation. Bottom: Layout of transmission lines from analog pixel down to digital pixel cell. Lines in red and blue intentionally emphasised to indicate lines with bigger spacing.}
\end{figure}

An extensive search for crosstalk has been carried out. No signs have been found except for a peculiar case tied to the arrangement of the transfer lines. In our telescope setup, events with 3~hits on the DUT have been studied. One of these hits must be compatible with a track fitted through the three reference planes. Such events occurred at a rate of a few percent for a typical choice of threshold where the three hits were spaced by an empty pixel. Within a column, the transfer lines are routed in groups of even and odd row addresses, hence the empty pixel in between the hits. The spacing of the lines is the same for most of the lines with a few exceptions. The missing entries for certain row addresses, see Fig.~\ref{fig:xtalk}, nicely match those cases with increased spacing. The crosstalk appears to originate from neighbouring transfer lines along the pixel column. To mitigate the effect, an adjustment to the line driver seems to be sufficient and will be implemented in the next version of the chip.
% TODO Mit Ivan Rücksprache halten ob wir da sogar mehr sagen können wie dass das bereits in Simulationen überprüft worden sei etc.

%==================================================
\section{High rate performance}
\label{sec:daqperf}
\begin{figure}[htbp]
\centering % \begin{center}/\end{center} takes some additional vertical space
\includegraphics[width=.7\textwidth]{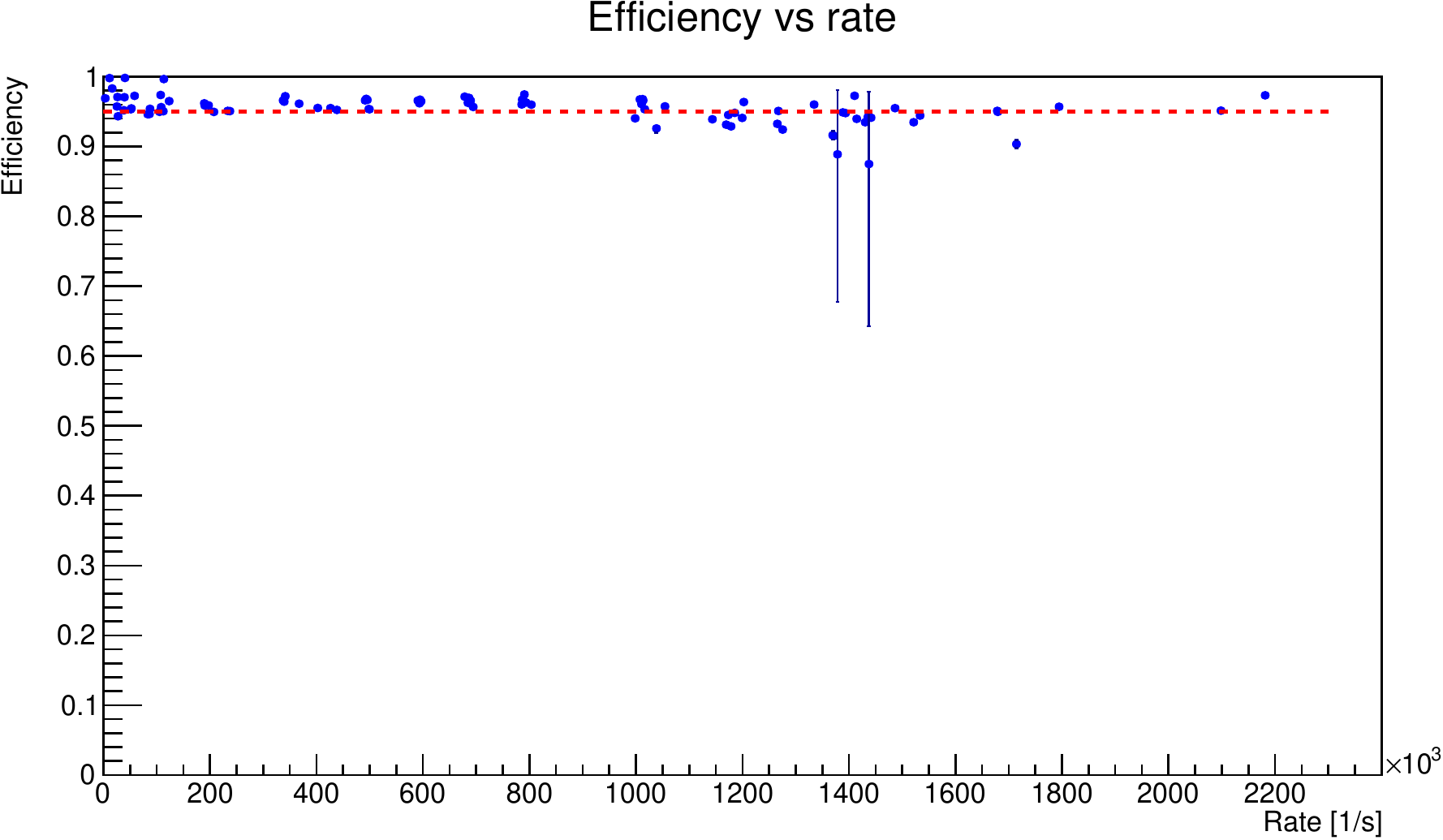} % new version of plot with better assumption for error bars and revisited efficiency correction, received Sep 29, 2016
\caption{\label{fig:effVsRate} Efficiency of the DUT plane at varying hit rates on a $5 \times 5$ pixel array. Horizontal line added at arbitrary value of 0.95 to guide the eye.}
\end{figure}

Rate dependent effects in the pixel cell have been tested in a high rate setup at MAMI. A beam of electrons at an energy of \SI{855}{\MeV} was focused on a sub-array of $5 \times 5$ pixels, keeping the overall hit rate well below any readout limitations. The analysis used a bigger time window of \SI{\pm 80}{\nano\second}. No rate dependency on the efficiency of the DUT has been observed for rates up to \SI{2.2e6}{\hertz / 25\,pixels}, which corresponds to \SI{1070}{\mega\hertz/\centi\metre\squared}, see Fig.~\ref{fig:effVsRate}. This is well above the rate of \SI{2.5}{\mega\hertz / \centi\metre\squared} required for the Mu3e experiment during phase~I. For comparison, the serialiser of the chip works at \SI{1.25}{\giga bits/\second}. One hit consumes \SI{40}{bits} (8b10b encoded including comma word), hence one data link is capable of handling at least \SI{30}{\mega hits/\second}. Or, a chip featuring the design target area of $\SI[parse-numbers = false]{20 \times 20}{\milli\metre\squared}$ will be limited to $\SI{7.8}{\mega hits / (\second \ensuremath{\cdot} \centi\metre\squared)}$ by one such link.

%==================================================
\section{Discussion and outlook}
\label{sec:discussion}
The MuPix7 chip has shown excellent performance, meeting requirements for the upcoming Mu3e experiment. In essence, at settings with a power consumptions of \SI{300}{\milli\watt/\centi\metre\squared}, an efficiency of better than 99.5\% and a time resolution of \SI{14.2}{\nano\second} have been measured. The sustainable data rate is way above the requirements.

While MuPix7 is a great success, it is still a test chip. The next version, MuPix8, is under way at time of writing. Among the changes are the enlargement of the column size to cover the target length of about \SI{20}{\milli\metre} (for use with Mu3e), mitigation and optimisation of the crosstalk, optimise pad layout for module integration studies, and the number of data links will be increased to 3. The latter will enable the chip to handle up to \SI{94}{\mega hits/\second} or about \SI{23}{\mega hits / (\second \ensuremath{\cdot} \centi\metre\squared)}.

%==================================================
\acknowledgments

We gratefully acknowledge the beamtimes provided by the following facilities: test beam facility at DESY Hamburg (Germany), a member of the Helmholtz Association (HGF), $\pi$M1 at Paul Scherrer Institut, Villigen (Switzerland), and MAMI at Institut für Kernphysik at the JGU Mainz (Germany).

S.~Dittmeier and L.~Huth acknowledge support by the \textit{International Max Planck Research School for Precision Tests of Fundamental Symmetries}. N.~Berger, A.~Kozlinskiy, I.~Sorokin, A.~Tyukin, and M.~Zimmerman acknowledge funding by the \emph{PRISMA cluster of excellence}, Mainz, Germany. N.~Berger, U.~Hartenstein, Q.~Huang, A.~Kozlinskiy, S.~Shrestha, D.~vom~Bruch, and F.~Wauters wish to thank Deutsche Forschungsgemeinschaft for support through the Emmy Noether program. H.~Augustin acknowledges support by the HighRR research training group (GRK 2058).



\begin{thebibliography}{99}

\bibitem{peric}
I. Perić, \emph{A novel monolithic pixelated particle detector implemented in high-voltage CMOS technology}, \emph{Nucl.Instrum.Meth.} {\bf A582} (2007) 876-885
\url{http://dx.doi.org/10.1016/j.nima.2007.07.115}

\bibitem{mpx7}
H. Augustin et. al., \emph{The MuPix System-on-Chip for the Mu3e Experiment},
arxiv:1603.08751,
\url{http://dx.doi.org/10.1016/j.nima.2016.06.095}

\bibitem{tele}
L. Huth, \emph{Development of a Tracking Telescope for Low Momentum Particles and High Rates consisting of HV-MAPS},
Master Thesis, Universität Heidelberg (2014).
\url{https://www.psi.ch/mu3e/ThesesEN/MasterHuth.pdf}

\bibitem{desytele}
\url{https://telescopes.desy.de/Main_Page}

% Please avoid comments such as "For a review'', "For some examples",
% "and references therein" or move them in the text. In general,
% please leave only references in the bibliography and move all
% accessory text in footnotes.

% Also, please have only one work for each \bibitem.


\end{thebibliography}
\end{document}